# Low-Cost and Detunable Wireless Resonator Glasses for Enhanced Eye MRI with Concurrent High-Quality Whole-Brain MRI

*Ming Lu [a*], Xiaoyue Yang [b*], Jason Moore [a,c], Pingping Li [a], Adam W. Anderson [a,d,e], John C. Gore [a,d,e], Seth A. Smith [a,d,e], Xinqiang Yan [a,b,e†]*

  a. Vanderbilt University Institute of Imaging Science, Vanderbilt University Medical Center, Nashville, TN, 37232, USA
  b. Department of Electrical and Computer Engineering, Vanderbilt University, Nashville, TN, 37232, USA
  c. Philips, Nashville, Tennessee, USA
  d. Department of Biomedical Engineering, Vanderbilt University, Nashville, TN, 37232, USA
  e. Department of Radiology and Radiological Sciences, Vanderbilt University Medical Center, Nashville, TN, 37232, USA

Xiaoyue Yang and Ming Lu contributed equally to this work.

**Corresponding Author (†):**

Xinqiang Yan, Ph.D.
Vanderbilt University Institute of Imaging Science
1161 21st Avenue South
Medical Center North, D-2205
Nashville, TN 37232-2310
Phone No: 1(615) 5253989
Email:    xinqiang.yan@vumc.org

**Submitted to MRM as a Technical Note**

**ABSTRACT:**

**Purpose:**

To develop and evaluate a wearable wireless resonator glasses design that enhances eye MRI signal-to-noise ratio (SNR) without compromising whole-brain image quality at 7 T.

**Methods:**

The device integrates two detunable LC loop resonators into a lightweight, 3D-printed frame positioned near the eyes. The resonators passively couple to a standard 2Tx/32Rx head coil without hardware modifications. Bench tests assessed tuning, isolation, and detuning performance. $B_1^+$ maps were measured in a head/shoulder phantom, and SNR maps were obtained in both phantom and in vivo experiments.

**Results:**

Bench measurements confirmed accurate tuning, strong inter-element isolation, and effective passive detuning. Phantom $B_1^+$ mapping showed negligible differences between configurations with and without the resonators. Phantom and in vivo imaging demonstrated up to a ~3-fold SNR gain in the eye region, with no measurable SNR loss in the brain.

**Conclusion:**

The wireless resonator glasses provide a low-cost, easy-to-use solution that improves ocular SNR while preserving whole-brain image quality, enabling both dedicated eye MRI and simultaneous eye–brain imaging at ultrahigh field.

**INTRODUCTION:**

Eye MRI plays a vital role in the non-invasive assessment of a wide range of ocular and neuro-ophthalmologic conditions, including optic neuropathies, ocular tumors, retinal abnormalities, and disorders of the visual pathway [1–8]. MRI offers exceptional soft tissue contrast, high spatial resolution, and the ability to capture both structural and functional information without ionizing radiation, making it particularly advantageous for longitudinal monitoring and pediatric imaging.

Despite these advantages, conventional MRI hardware imposes significant limitations on image quality in the ocular region when standard head coils are used [9–12]. Head coils are designed with a large inner diameter (especially in the anterior/posterior direction) to accommodate a broad range of adult head sizes and ensure patient comfort. As a result, in most situations, the eyes are positioned several centimeters away from the nearest receive elements, leading to a substantial reduction in signal-to-noise ratio (SNR) in the orbital region. Furthermore, the loop elements in head coils are typically much larger than the optimal size for ocular imaging, further limiting SNR performance.

Although dedicated eye coils can significantly improve image quality by placing appropriately sized receive elements in close proximity to the eyes, they are not part of the standard configuration on research or clinical MRI systems [13–15]. These dedicated eye coils are either built in-house or purchased separately. They are often prohibitively expensive and not even available at ultrahigh fields. Even when available, dedicated eye coils are typically limited in field of view and solely for the orbit [13,15], thereby restricting their utility in studies involving the visual pathways or brain.

Meanwhile, numerous structural MRI studies have demonstrated that ocular diseases such as glaucoma, amblyopia, macular degeneration, and hereditary retinal dystrophies are often associated with changes in the optic nerve, visual pathways, and even cortical regions [7,16–25]. These findings underscore the need for imaging solutions that not only enhance SNR in the eye but also preserve the ability to perform high-quality whole-brain MRI within the same setup.

To address these challenges, we propose a novel pair of glasses frames with integrated wireless resonators to enhance SNR in eye MRI. This device consists of detunable, passive resonant structures that couple inductively to existing head coils without requiring any physical connection or hardware modification. They can be easily positioned near the eyes, significantly reducing the distance between the resonant elements and the region of interest. Its detunable design ensures safe operation during RF transmission and compatibility with various head receive arrays. Most importantly, it improves the SNR in the ocular region while maintaining similar SNR performance in the brain.

We demonstrate the feasibility and efficacy of this approach through bench test, as well as phantom and in vivo human imaging experiments. The results show substantial improvements in orbital SNR, with no measurable SNR degradation in the brain. This wireless resonator glasses provides a low-cost, and easy-to-deploy solution for enhancing eye MRI while enabling concurrent high-quality brain imaging, potentially expanding clinical access without requiring dedicated hardware.

**METHODS:**

**Hardware design and fabrication**

Figure 1A shows the circuit schematic of the wireless resonator array (hereafter referred to as <u>wireless resonator glasses</u>) designed for eye MRI at 7 T. It consists of two LC loop resonators, each precisely tuned to the Larmor frequency (298 MHz) of our 7 T whole-body scanner (Philips, Best, Netherland). Each resonator contains three distributed capacitors (labeled $C_t$ in Figure 1A) and is equipped with a passive detuning circuit to ensure automatic deactivation during the transmit (Tx) period. A bridge capacitive network (labeled $C_d$ in Figure 1A) is used to eliminate mutual coupling between the two loops [26]. For added safety, MRI-compatible fuses are integrated into each loop to prevent unintended resonant behavior in the event of detuning failure.

This wireless resonator glasses design operates on the same principle as in previous studies [27–32]: it works in conjunction with the local array during the receive (Rx) period and remains effectively invisible using passive detune circuit consists of lumped components during the Tx period. At 7 T, the local array is the Nova volume Tx/32-channel Rx head coil (Nova Medical, Wilmington, MA, USA), a widely used standard configuration. During the Rx period, the resonators couple passively to the 32-channel receive array, enhancing local signal reception through close-proximity inductive coupling without requiring any direct electrical connection or active circuitry. Importantly, the design does not interfere with the volume transmit coil during RF transmission. Consequently, no adjustments to transmit-related parameters, such as reference power, voltage, or specific absorption rate (SAR), are necessary. The reason for making the wireless resonators completely invisible or fully detuned, rather than only partially detuned [33], is that partly because such a design does not require Tx-related parameter adjustment and partly because the eye region is already covered by the Nova 2-channel Tx coil and does not require additional $B_1^+$ field enhancement from the wireless resonators.

The two resonators, including tin-coated wires and components, were embedded within a customized glasses frame, as shown in Figure 1B. The frame was designed in SolidWorks (Dassault Systèmes SE, Vélizy-Villacoublay, France) and 3D-printed using a Formlabs 3 printer (Formlabs Inc., Somerville, MA, USA). This wearable design positions the resonators in close proximity to the eyes while maintaining patient comfort. The eye region was deliberately left open to ensure compatibility with eye-tracking systems during MRI.

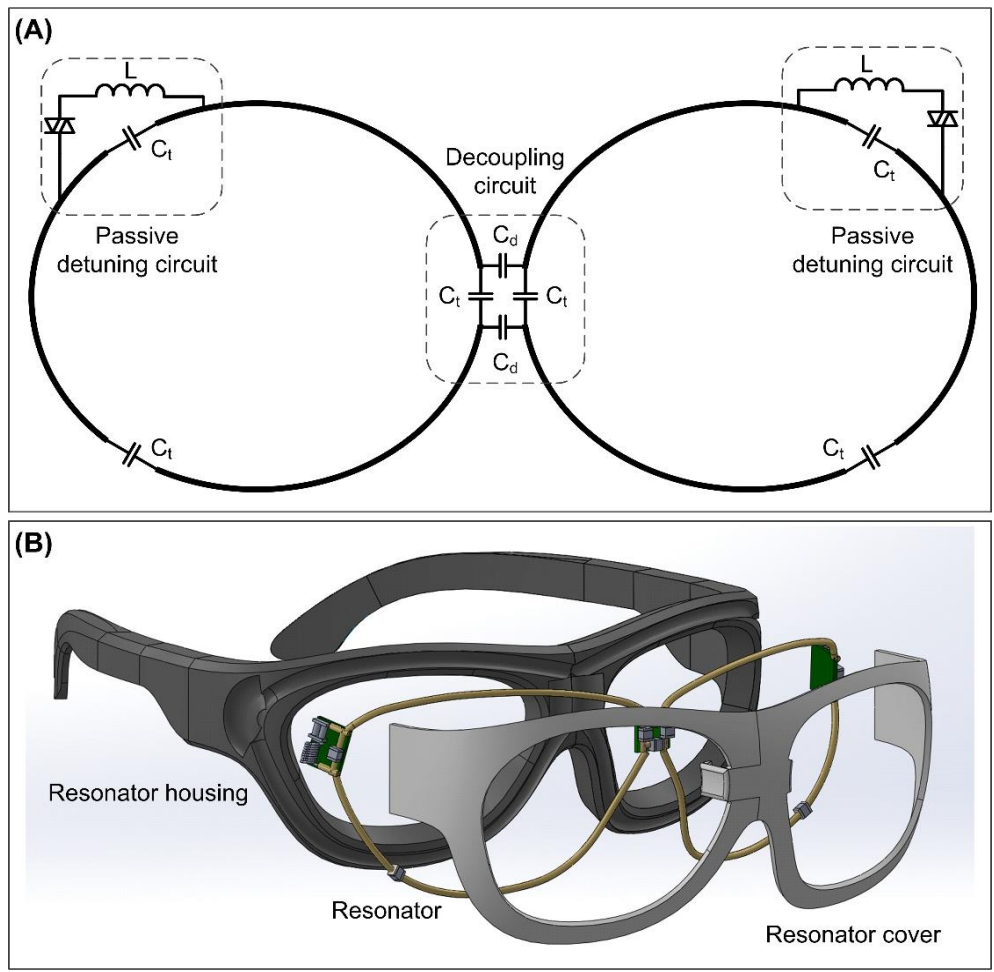

**Figure 1:** Circuit diagram (A) and CAD design (B) of the wireless resonator glasses.

## Bench test

Bench testing was conducted to ensure that both wireless resonators were properly tuned to the Larmor frequency, were well decoupled, and also can be effectively detuned. This test was performed using a pair of well-decoupled double pick-up probes. The decoupling performance was assessed by comparing the unloaded quality ($Q_{un}$) factor of the two-resonator system to that of a single ideal resonator, ensuring the merged resonance peak resembled that of a single resonator. Passive detuning performance was also evaluated on the bench by actively turning the cross diodes ON and OFF. While this active switching does not fully represent in-scan conditions where the diodes are switched passively, it provides a resonant-based validation of the detuning effectiveness when the diodes are turned ON during the Tx period.

**MRI experiments**

MRI experiments were conducted at 7T with the Nova volume Tx/32-channe Rx head coil alone, and in combination with the wireless resonator glasses.

We first investigated $B_1^+$ experiments using a head/shoulder-shaped phantom. The phantom was fabricated with distilled water, sugar, NaCl, and gel, and was designed to mimic human tissues at 298 MHz, with a measured conductivity of approximately 0.5 S/m and a relative permittivity of ~55. The $B_1^+$ maps were measured using the TurboFLASH method [34], which was optimized for the ultrahigh field MRI. The imaging parameters included a field of view (FOV) of 250 × 250 mm², a slice thickness of 3 mm, and an in-plane resolution of 3 × 3 mm².

We then assessed the SNR on the head/shoulder gel phantom. SNR maps were calculated from gradient recalled echo (GRE) images with the following parameters: axial orientation, FOV = 250 × 250 mm², TR/TE = 1000/1.96 ms, nominal flip angle (FA) = 70°, in-plane resolution = 1 × 1 mm², slice thickness = 5 mm, and number of averages = 1. Besides GRE images with a FA of 70°, noise-only maps were acquired using the exact same parameter, except with the RF power turned off. The noise correlation matrix of the 32-channel Rx array was calculated based on the noise-only data, and SNR maps with the optimal combination method were generated based on the GRE images and noise data [35].

In addition to phantom images, T1-weighted (T1W) and T2-weighted (T2W) images of a healthy volunteer were acquired using the Nova 2Tx/32Rx coil, without and with the wireless resonator glasses. SNR maps of the human images were calculated based on the scanner's default reconstructed images. The imaging protocols were as follows:

(1) T1W imaging: Axial acquisition, TR/TE = 5.0/2.2 ms, FA = 7°, FOV = 220 × 220 × 40 mm³, voxel size = 1.0 × 1.0 × 1.0 mm³, bandwidth (BW) = 505 Hz/pixel, number of averages = 1;

(2) T2W: axial slices, TSE, TR/TE = 3000/302 ms, FA/refocusing FA = 100/35°, FOV = 2770 × 100 × 34 mm³, voxel size = 0.7 × 0.7 × 0.7 mm³, BW = 1157 Hz/pixel, number of averages = 1.

Safety tests for gradient-induced and RF-induced heating were conducted before human imaging. All experimental procedures were approved by the local institutional review board (IRB #060730), and participants provided informed written consent.

**RESULTS:**

**<u>Bench test results</u>**

Figures 2B-C shows the measured $S_{21}$ plots and calculated Q-factors of the double-pickup probes when both resonators are well-tuned to 298 MHz and are well-decoupled, without any loading. The double-pickup probe was initially adjusted to achieve high isolation, maintaining <-70 dB even without the presence of resonators (Figure 2A). The average $Q_{un}$ of these two resonators is 292.8. Additionally, we present the $S_{21}$ plot of a single ideal resonator (without the presence of another resonator), which exhibits a $Q_{un}$ of 313.3 (Figure 2D). The distinct single-resonator peak observed in the two-resonator array, along with the high $Q_{un}$ value (93.5% of the $Q_{un}$ of a single ideal resonator), demonstrates the excellent isolation and decoupling between the two resonators.

Figure 2E shows the $S_{21}$ plot of the pickup probes and the calculated $Q_L$ of the right resonator when loaded with the head/shoulder phantom. Under phantom loading, the $Q_L$ is 32.3, corresponding to a high Q-ratio of approximately 9.1 for this wireless resonator. The left resonator exhibited a similarly high Q-ratio of ~9, though its data are not shown here. Figure 2F presents the $S_{21}$ plots when the cross diodes are actively turned ON. A significant reduction in the $S_{21}$ magnitude at 298 MHz (from –30.4 dB to –61.2 dB) demonstrates the effective detuning capability of the resonator when the cross diodes are activated during RF transmission.

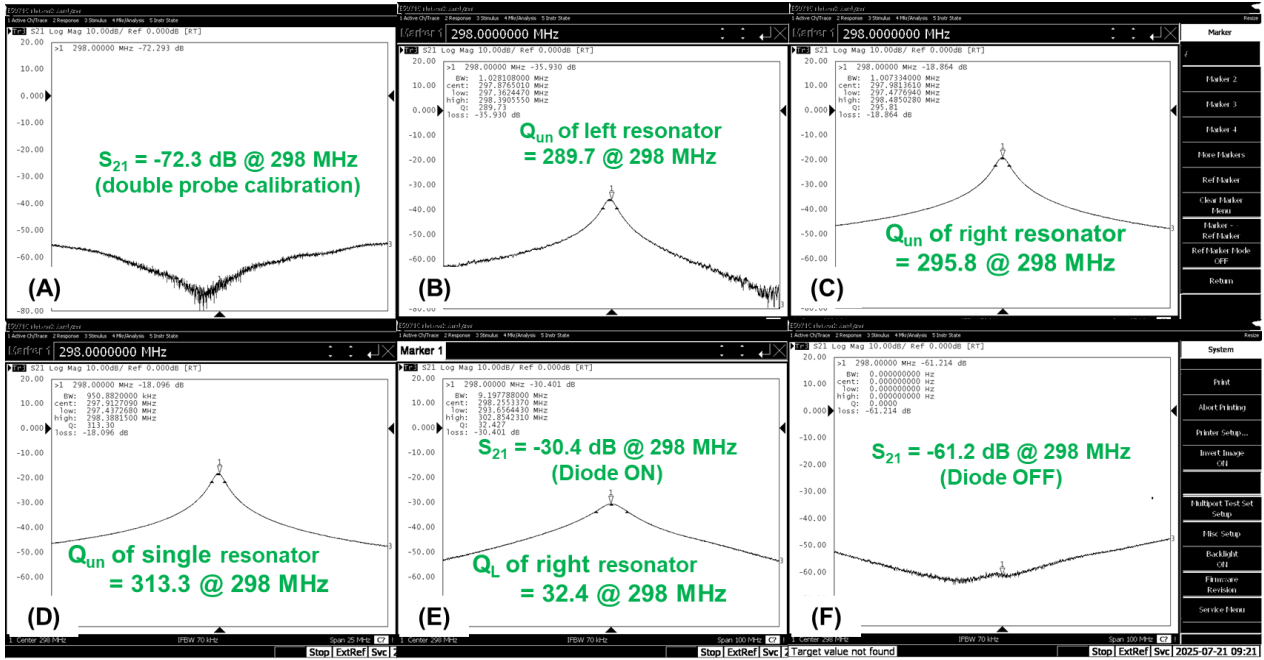

**Figure 2** Bench test results under different scenarios using double pickup probes. (A) $S_{21}$ measurement of the double probes without resonators or loadings, achieving a value of <-70 dB to ensure the double probes are well-decoupled. (B) $Q_{un}$ of the left resonator in the wireless resonator glasses. (C) $Q_{un}$ of the right resonator. (D) Baseline comparison of the $Q_{un}$ of a single resonator, measured without the presence of the other resonator. (E) $Q_L$ of the right resonator. (F) $S_{21}$ measurement of the double probes on the right resonator with the cross diode turned OFF.

## $B_1^+$ results

Figure 3 shows the measured axial $B_1^+$ maps of the head/shoulder phantom acquired using the Nova coil under identical input power for two configurations: without and with the wireless resonator glasses. The slice shown was selected to approximately intersect the center of the wireless resonator glasses (top row in Figure 3). Complete multi-slice $B_1^+$ mapping results from the same experiment are presented in Supporting Figure S1. Under identical input power, the average flip angles across the head-shaped phantom differed by less than 5% between the two configurations. This outcome aligns with expectations, as the wireless resonators are detunable and remain "invisible" during RF transmission, producing negligible changes in the $B_1^+$ field.

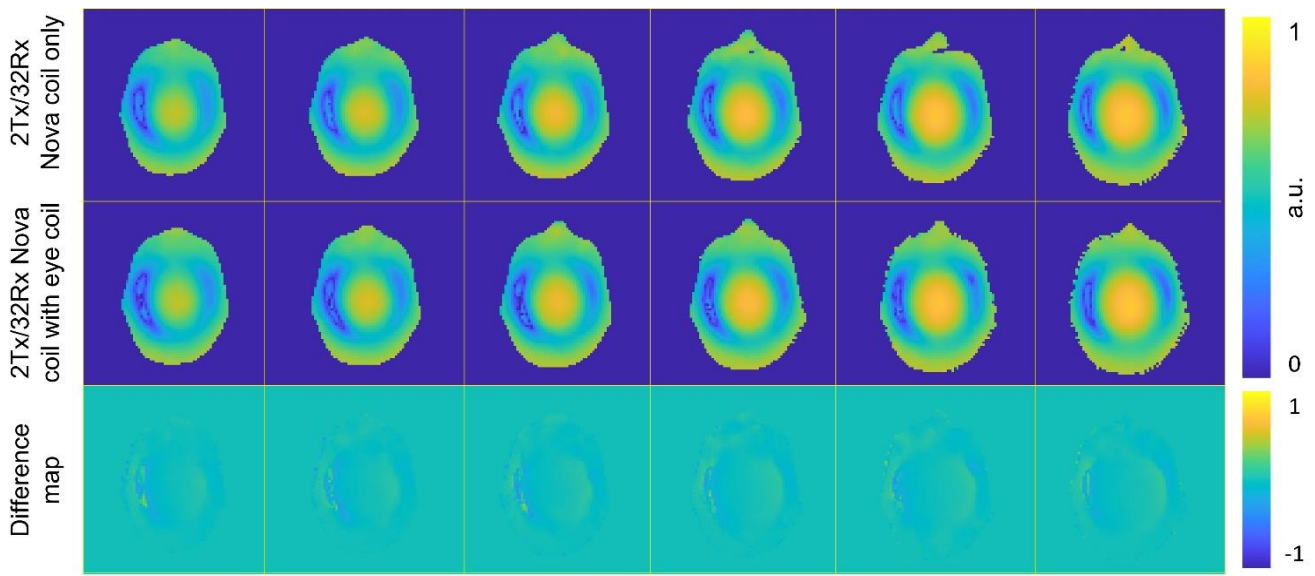

**Figure 3:** Measured axial $B_1^+$ maps of the head/shoulder phantom without (left column) and with (right column) the wireless resonator glasses.

## **Phantom and in vivo SNR**

Figure 4A shows the measured SNR maps in the axial slice using the Nova 2Tx/32Rx coil, without and with the wireless resonator glasses. On average, a 2.6-fold increase in SNR was observed in the eye region, as indicated by the elliptical circle in the figure. Meanwhile, the SNR in other areas remained at the same level, demonstrating that the wireless resonator can maintain same level of SNR in the brain area while significantly improving SNR in the eye region. Additionally, a 1D plot of the SNR map was generated (Figure 4B), showing that the SNR benefits are present at depths less than 5 cm but diminish beyond 5 cm in the phantom.

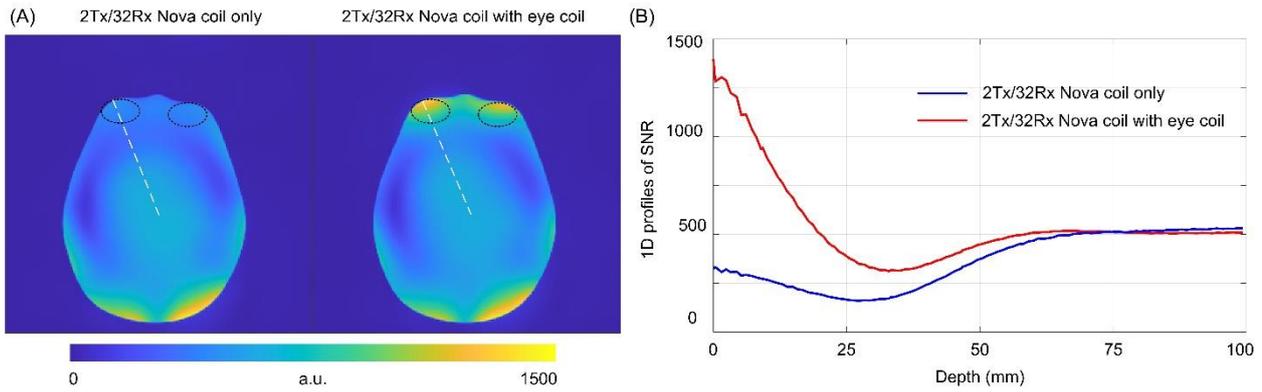

**Figure 4:** (A) Measured SNR maps in the axial slice without and with the wireless resonator glasses. (B) 1D SNR plots along the dotted lines shown in Figure 4A.

Figure 5A shows the in vivo T1-weighted whole-brain images acquired with the Nova 2Tx/32Rx coil, without and with the wireless resonator glasses. Figure 5B presents representative T2-weighted images focused on the eye region. Consistent with the phantom results, the addition of the resonator produced a localized SNR gain, most pronounced in the eye region, with an average 3.2-fold increase in SNR. Because of the large SNR differences, separate color bars were used for the T2-weighted SNR maps with and without the resonators. Importantly, SNR across most of the brain remained unchanged, confirming that the resonator does not compromise overall brain image quality. Furthermore, the frontal lobe also exhibited improved SNR with wireless resonator glasses, likely due to its anatomical proximity to the eyes and resonators.

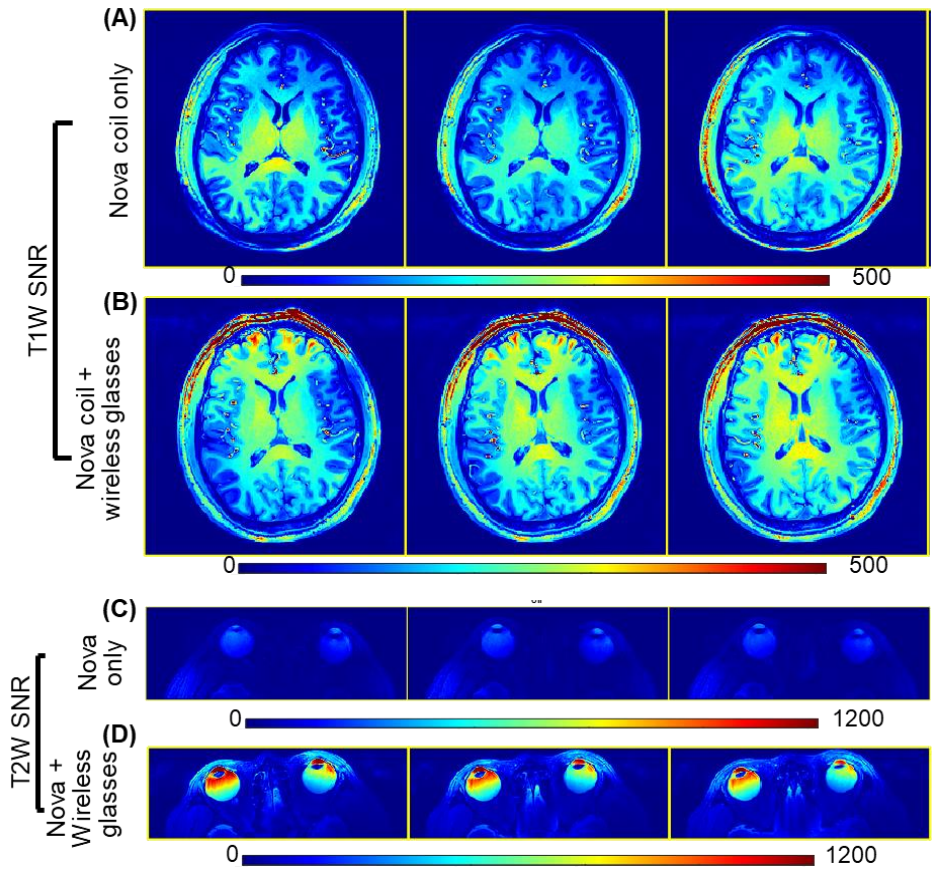

**Figure 5:** In vivo SNR maps with the Nova 2Tx/32Rx coil based on T1W (A) and T2W (B) images, without and with the wireless resonator glasses. The resonator yielded a ~3-fold SNR increase in the eyes and improved SNR in the frontal lobe, while preserving overall brain image quality.

**DISCUSSIONS and CONCLUSION**

This study demonstrates the feasibility and effectiveness of a wearable wireless resonator frame for enhancing SNR in the eye region during MRI without compromising brain image quality. By embedding detunable LC loop resonators into a lightweight, 3D-printed frame positioned close to the eyes, we achieved up to a ~3-fold SNR improvement in the eye region in phantom and in vivo experiments. Importantly, the brain SNR remained unchanged mostly in this setup.

The wireless resonator glasses effectively enhance eye imaging at 7T MRI, leveraging near-field sensitivity for high SNR close to the resonator surface. The versatile design principles can be adapted for imaging other superficial structures, such as the ears, temporomandibular joints, or cortical regions, by adjusting loop size, shape, or positioning. While optimized for 7T, the approach can be extended to lower field strengths like 1.5T or 3T, with adjustments for differences in SNR and penetration depth, enabling broader applications in MRI.

From a practical standpoint, the fully 3D-printed frame offers a customizable, low-cost, and patient-friendly solution. The frame's mechanical flexibility allows it to accommodate a range of head sizes, while an optional adjustable elastic band ensures stable positioning during scanning. The open-eye design preserves compatibility with eye-tracking systems, which is valuable for functional MRI studies involving visual stimuli. Furthermore, the passive, inductively coupled configuration allows seamless integration with existing head coils without hardware modifications, changes to Tx-related parameters (e.g., reference scaling, SAR settings), or adjustments to the scanning workflow.

Compared with dedicated wired eye coils, the wearable wireless resonator frame offers multiple advantages. While dedicated eye coils can improve ocular SNR, they are typically expensive, structurally complex, and restricted to orbital imaging—lacking the capability for whole-brain imaging within the same scan session. In contrast, the wearable wireless resonator frame not only enhances ocular SNR while preserving full-brain imaging capability, but also has a much lower fabrication cost, a simpler structure, and a straightforward setup process. These factors enable broader accessibility, faster patient preparation, and easier adoption in both research and clinical environments.

It is worth noting that conventional dedicated eye coils, when used as receive-only devices, typically incorporate active detuning circuits that can be explicitly triggered to disable the coil during transmission. In contrast, the wireless resonator frame relies on passive detuning. However, our measured $B_1^+$ maps confirmed that the passive detuning in our design is sufficient in practice, with the resonators effectively "invisible" during the Tx phase. Additionally, each resonator loop incorporates an MRI-compatible fuse to prevent unintended resonant behavior in the unlikely event of detuning failure.

This work demonstrates promising advancements in wireless resonator design for ocular imaging, but several limitations and areas for improvement should be noted. First, the wearable glasses design adds minimal additional space within the coil. However, the 7T Nova receive coil itself is relatively compact, and fitting can be challenging for subjects with very large heads. In this study, the head circumference was 60 cm, which represents a large head size (above the 95th percentile for men and well above average for women). We found that this was close to the upper limit of what could be accommodated by the coil. For individuals with larger head sizes, wearing wireless resonator glasses inside the coil may not be feasible. Second, the depth-dependent SNR profile (Figure 4) indicates pronounced enhancement within approximately 5 cm of the resonators, with diminishing gains at greater depths, a characteristic of near-field sensitivity in loop-based receive structures [36]. To address deeper targets, such as the intracranial portion of the optic nerve, larger resonators or an increased number of resonators may be necessary. Third, the clinical value has not been extensively demonstrated. Although higher SNR typically translates to improved resolution and/or shorter scan times, further validation in larger patient cohorts is essential to establish the device's practical utility in clinical applications. Such studies would provide stronger evidence of its impact on diagnostic accuracy and patient outcomes.

**ACKNOWLEDGEMENTS:**

Ming Lu and Xiaoyue Yang contributed equally to this work. This work was supported in part by NIH grants R01 EB031078, R21 EB029639, R21 EB037763, R03 EB034366 and S10 OD030389. The content is solely the responsibility of the authors and does not necessarily represent the official views of the NIH.